# Designing and Evaluating an AI-enhanced Immersive Multidisciplinary Simulation (AIMS) for Interprofessional Education

Ruijie Wang[1] [0009-0004-7196-4410] and *Jie Lu[2] [0000-0002-7466-6177] and Bo Pei[3] [0000-0002-6328-6929] and Evonne Jones[2] [0009-0001-0455-8533] and James Brinson[4] and Timothy Brown[2] [0000-0002-6470-7406]
[1]University of Florida, Gainesville, FL 32611, USA
[2]University of Georgia, Athens, GA 30602
[3]University of South Florida, Tampa, FL 33620
[4]Augusta University, Augusta, GA 30912
jlu@uga.edu

**Abstract.** Interprofessional education has long relied on case studies and the use of standardized patients to support teamwork, communication, and related collaborative competencies among healthcare professionals. However, traditional approaches are often limited by cost, scalability, and inability to mimic the dynamic complexity of real-world clinical scenarios. To address these challenges, we designed and developed AIMS (AI-enhanced Immersive Multidisciplinary Simulations), a virtual simulation that integrates a large language model (Gemini-2.5-Flash), a Unity-based virtual environment engine, and a character creation pipeline to support synchronized, multimodal interactions between the user and the virtual patient. AIMS was designed to enhance collaborative clinical reasoning and health promotion competencies among students from pharmacy, medicine, nursing, and social work. A formal usability testing session was conducted which participants assumed professional roles on a healthcare team and engaged in a mix of scripted and unscripted conversations. Participants explored the patient's symptoms, social context, and care needs. Usability issues were identified (e.g., audio routing, response latency) and used to guide subsequent refinements. Findings in general suggest that AIMS supports realistic, profession-specific and contextually appropriate conversations. We discussed technical innovations of AIMS and concluded with future directions.

## 1 Introduction

Effective education of healthcare professionals (HCPs) is among the most critical endeavors of our society, as we rely on the expertise of physicians, pharmacists, nurses, and allied health practitioners to sustain the health of individuals and communities. Interprofessional education (IPE), in which students from multiple healthcare professions learn to work as a team, has been widely recognized as an essential approach for preparing a collaborative practice-ready workforce [1] and improving patient health outcomes [2].

Traditionally, IPE has relied on descriptive case studies [3] and live simulations with standardized patients (SPs) as primary methods to facilitate the development of key Interprofessional Education Collaborative (IPEC) competencies, such as roles and responsibilities, communication, teamwork, along with values and ethics [4]. While these approaches provide valuable learning and training experiences, they also pose significant challenges in effectively preparing students for the complexities of real-world clinical practice. Case studies, although cost-effective, are often limited to theoretical and hypothetical discussions that fail to reproduce the dynamic and unpredictable nature of clinical environments. Live simulations with SPs can offer authentic interactions, but incur high costs associated with hiring and training actors [5], therefore limiting opportunities for repeated practice. These limitations highlight an urgent need and call for innovative training solutions that are realistic, dynamic, scalable, and cost-effective. Recent advances in artificial intelligence (AI) and immersive technology, such as virtual simulations, have demonstrated promises for transforming interprofessional training. For instance, [6] facilitated effective team training among students from two healthcare professions using virtual simulations. Similarly, multiple meta-analyses and systematic reviews [7–9] have found that immersive experiences can enhance post-intervention knowledge retention in health professions education compared to traditional learning methods. Leveraging AI and simulation techniques, we developed AIMS (AI-enhanced Immersive Multidisciplinary Simulations), a collaborative screen-based virtual simulation [10] that integrates an AI-driven virtual patient into two distinct clinical settings. Hosted at a major research university in the southeastern United States and in partnership with two public institutions on the East Coast, AIMS enables healthcare provider students to engage in dynamic, authentic clinical encounters where the virtual patient responds in real-time to student prompts. This paper focuses and reports on the design, development, and early evaluation of AIMS as a potential alternative to SP-based training that supports the development and practice of core IPEC competencies through high-fidelity interactions. Unlike VR-heavy approaches, our innovation lies in implementing conversational AI, role-based logic, and synchronized embodied responses in a screen-based environment that is accessible, cost-effective, and deployable across institutions.

This paper is structured as follows: Section 2 reviews related work on virtual simulation applications and the use of conversational AI in healthcare profession education. Commonly reported challenges in design and implementation were discussed. Section 3 provides the institutional context and program description of AIMS, followed by a detailed report on methods of the design, development, and evaluation. Results of a preliminary user testing session are presented in Section 4, which record usability issues rated based on individual severity. The paper concludes with a discussion and directions for future work.

## 2 Related Work

### 2.1 Immersive Technology and Conversational AI for Teaching and Training in Healthcare Education

Simulation-based technologies have been widely adopted in healthcare education to improve students' practical skills, knowledge development, and patient safety [11–13]. Early work in this domain explored interactive virtual humans capable of spoken dialogue to augment clinical skills training. For example, researchers [14] developed a virtual patient environment in Second Life where students can engage in natural conversations with AI-controlled avatars in clinical scenarios. It was created to provide a scalable, immersive, and cost-alternative to standardized patient programs that allows learners to practice clinical skills in a risk-free setting. Al-dahir and colleagues [15] developed an interactive, online virtual patient simulation using a branched learning platform (DecisionSimulation/VpSIM) that mimicked a hospital case of atrial fibrillation. This simulation allowed students to navigate patient care decisions through dynamic, looped scenarios where choices led to different clinical outcomes. In parallel, foundational work by Kenny and colleagues [16] conceptualized and developed intelligent virtual patients, which were embodied conversational agents that can simulate believable clinical interactions with learners to practice interviewing skills, symptom assessment, and clinical reasoning in a controlled setting. Building on this work, Bickmore's [17] research further demonstrated how animated virtual agents can support health communication and counseling training by engaging learners in repeated, adaptive conversations that reinforce motivational and educational objectives. More practically oriented simulation programs have also been integrated into curricula, for instance, researchers [18] developed a computer-assisted simulation for newborn evaluation and care to train fourth-year midwifery students on deliver-room procedures such as delivery room preparation, initial neonatal assessment, and safety protocols. Considering the increasing adoption and implementation of simulations in the field, [19] designed a Conceptual Framework for Simulation in Healthcare Education to support the pedagogically sound development of simulation-based learning, aiming to improve instructional design, educator training, and alignment between simulation activities and learning outcomes.

### 2.2 Evidence of Immersive Technology and Conversational AI Effectiveness in Healthcare Education

Simulation-based training has demonstrated significant potential to enhance learning across various health professions. In a review study by Elendu and colleagues [20], simulation-based training can promote healthcare professionals' learning experience and clinical competency. Findings suggested that the incorporation of simulation into medical education may enhance both learning outcomes and accessibility. [21] reported that simulation-based learning significantly improves medical student performance by an average of 18%, with first-year students benefiting more (22% improvement). More specifically, a quasi-experimental study [18] reported that learners in the simulation group demonstrated significantly higher performance in newborn evaluation and care skills compared with a control group, supporting the effectiveness of this simulation-based instructional design in clinical education. In a systematic review [22], identified virtual patients as an increasingly important tool in modern medical education for developing clinical reasoning skills and decision-making capabilities, with the potential to emerge as the primary method for creating interactive patient scenarios. Collectively, these studies demonstrate immersive simulation applications' capacity to deliver authentic, safe, and repeatable practice opportunities that can ultimately maximize educational and clinical outcomes.

### 2.3 Challenges and Research Goals

The use of virtual simulations and 360 videos in IPE has shown to support the development of interprofessional competencies, including team communication and collaborative clinical reasoning [23, 24]. However, despite its promise, several studies have noted persistent challenges in adopting these technologies at scale. Commonly reported issues include technical glitches (e.g., latency), cybersickness, and the loss of face-to-face communication inherent in virtual environments [25, 26]. To address these limitations, we developed AIMS, an AI-driven, multimodal virtual simulation designed to closely replicate authentic clinical scenarios while supporting dynamic, real-time interactions between users and the virtual patient.

## 3 Context and Program Description

The annual IPE event is an online session hosted via Zoom by our home institution and joined collaboratively by two other institutions in the United States. It is designed to introduce healthcare provider students to core competencies in teamwork, communication, and understanding of roles and responsibilities. To facilitate robust learning experiences, second-year medical students lead teams that consists of third-year pharmacy students, undergraduate or graduate social work students, and second-year Doctor of Nursing Practice students. This year's event will include 40 teams, each comprising ideally two students from each profession. Each team will be facilitated by a trained faculty member. Traditionally, the event uses a written case study to prompt discussion and role modeling. While effective for theoretical exploration, this format does not allow for direct patient interaction, which is a critical component for developing practical skills in patient interviewing and collaborative care planning.

To address this limitation, AIMS presents two immersive and interactive clinical scenarios: an Emergency Department (ED) counter and a primary care office visit. Central to this innovation is the creation of a sophisticated virtual patient avatar, Jane Ryan, a 38-year old white female presenting to the ED with an acute Urinary Tract Infection (UTI), later diagnosed with Opioid Use Disorder (OUD). Leveraging the Gemini-2.5-Flash multimodal application programming interface (API), AIMS allows healthcare provider students from each team to take turns interacting with Ms. Ryan and engaging in dynamic, realistic conversations that gradually unfold the case. Faculty facilitators guide the discussion as each team follows the patient's care journey, from the ED to the referred primary care office visit and eventually to a medication-assisted treatment program. AIMS is hosted on the university's server and can be accessed via web browser from personal computers or laptops, enhancing both flexibility and accessibility for remote learning.

## 4 Method

### 4.1 Technical Architecture

AIMS presents two clinical environments, an ED and a primary care office, each representing a distinct phase of the patient's care journey. Within each setting, students interact with the virtual patient from a first-person viewpoint and observe a character model capable of switching poses (e.g., sitting upright on an exam bed, lying down) to reflect the evolving clinical context.

As shown in Fig. 1, AIMS is built upon and facilitated by three interconnected engines: a Character Creation Engine, a Multimodal AI Engine, and a Virtual Environment engine. Each engine plays a distinct role in enabling immersive, responsive simulation-based learning across clinical contexts.

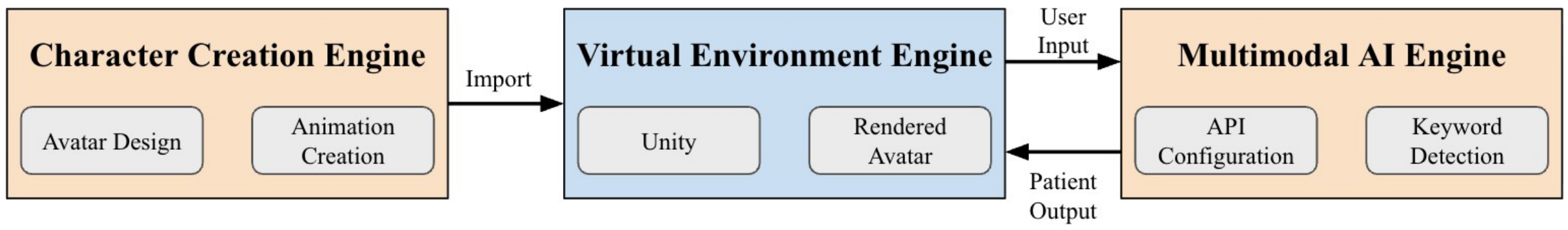

**Fig. 1.** The technical architecture of AIMS

### 4.2 Character Creation Engine

The virtual patient was first created in Reallusion [27] and then integrated into each clinical environments built using Unity [28]. Initial avatar configurations such as appearances, fashion styles, and default postures were developed with the subject matter expert (SME) to ensure clinical authenticity and visual fidelity for both scenarios. Dynamic physical gestures were scripted with appropriate facial expressions, with each animation clip programmed and activated by specific keywords (1). These design elements significantly enhance the clinical realism of the simulation by bringing in emotional nuance and embodied presence, allowing students from various healthcare professions to practice in environments that closely mirror real-world patient encounters.

**Table 1.** Unique keywords map to animations in Scene 1

| **Keywords** | **Animations** |
|---|---|
| "fever", "chills" | Patient shakes head |
| "symptoms", "discharge" | Patient lowers her head when answering awkward questions. Looks embarrassed. |
| "have sex", "husband", "kids" | A slight smile on patient's face |
| "have any questions?" | Looks confused (mainly the facial expression) |
| General conversation | A slight movement of one hand, a natural body gesture as if talking to someone while lying in bed with the upper body being upright. Facial expression remains natural. |

**4.2.1 Scene 1: Emergency Department.** The patient avatar in this scene is depicted as a young woman in a long-sleeved shirt, positioned sitting upright in a hospital bed with her head propping up. We imported a prefab bed into Reallusion's Character Creator (CC), adjusted the headrest angle and mesh alignment so that the avatar can recline while the upper body maintaining an upright position. We finished up the static model by adjusting the avatar's spinal and leg joints to match the bed's geometry, getting it ready for animation creation.

Animations were developed in iClone Editor, which provides preset options (e.g., speaking, confused talking) that we further refined to match specific conversational scenarios. The animation workflow was frame-based and we used iClone to generate transition clips by inserting keyframes. As shown in Table 1, the head-shake animation was associated with any mentioning of "fever" or "chills" in user inquires that would trigger a negative response, reflecting how humans naturally shake their heads with verbal rejections. The head-lowering animation was designed to restrict downward motion to the initial frames, followed by a gradual upward lift. Facial expressions such as smiling or confusion were driven by specific keyword triggers while preserving the character's baseline speaking gestures (e.g., lip movement, blinking). These intentional design choices help maintain continuity in overall motion, creating human-like and authentic interactions.

Fig. 2 shows the "confusion" posture in a static stage, in which the avatar slightly leans forward with her arms extended, accompanied with facial furrow that reinforces the emotion. This example illustrates the design approach for the confusion animation during conversational interactions.

**4.2.2 Scene 2: Primary Care Office.** We applied the same avatar and animation creation pipeline to Scene 2. The patient avatar remains the same woman but with different outfit. In this scene, Ms. Ryan's initial static posture was adjusted to sit fully upright rather than reclined, simulating a typical human behavior in a primary care context. Fig. 3 shows the editing view in iClone Editor prior to exporting the character to Unity.

The animation design for Scene 2 consists of three states, with one of them an idle state in which the character sits still and occasionally looks around. A "feel reluctant" animation shows frowning on the avatar's face with her hand touching the head occasionally. A "feel ashamed" animation has the avatar lowering her head during speech before slowly raising it again, conveying a sense of discomfort and embarrassment when the conversation unfolds around the discussion of her Vicodin use. As in Scene 1, these animations were created using keyframes, tested with sample interactions to refine timing, and exported as individual clips for integration into Unity. Table 2 documents how the keywords are mapped to animations in this scene.

### 4.3 Multimodal AI Engine

The Multimodal AI Engine is powered using Gemini-2.5-Flash to support real-time, speech-based dialogues between the virtual patient and healthcare provider students. A detailed, scripted clinical narrative was developed by the SME, which was used to train the API while serving as the knowledge base. System behavior was configured and iteratively refined using prompt engineering techniques. For example, we defined core conversational guidelines, response strategies, trigger words tied to each professional role, and contextual role detection mechanisms to ensure response relevance. The patient avatar's speaking style was configured explicitly with constraints such as "speak with conviction about your experiences" and "do not reveal Vicodin use when asked about medication history the first time". Audio input from students is captured through the Virtual Environment Engine and processed by the API, which generates and returns contextually appropriate verbal responses as output.

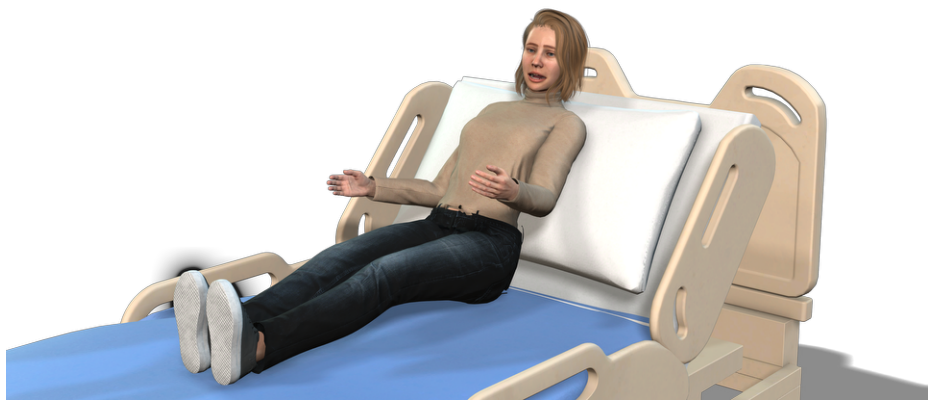

**Fig. 2.** The patient expressing confusion with a frowning facial expression

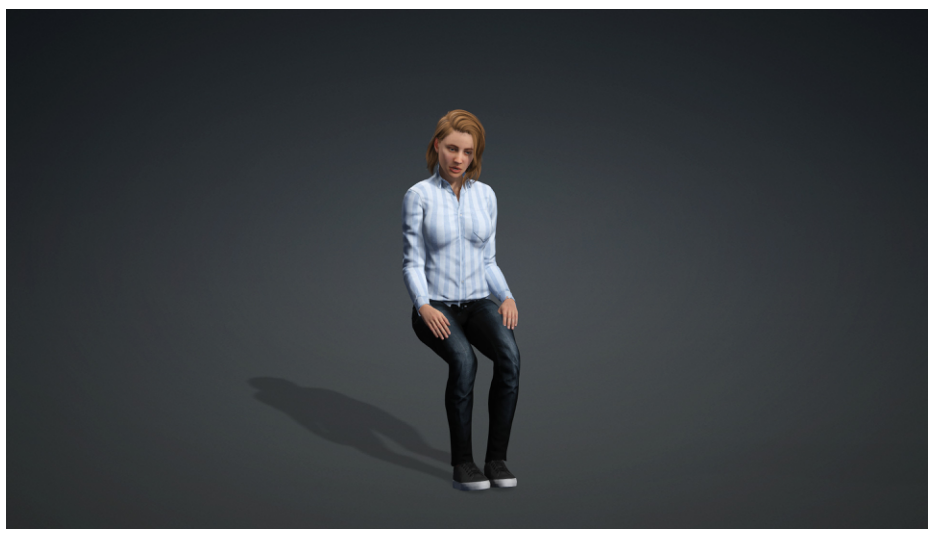

**Fig. 3.** The patient looking ashamed/embarrassed when answering awkward questions

**Table 2.** Unique keywords map to animations in Scene 2

| Keywords | Animations |
|---|---|
| "What brought you in?", "How can I help you?" | Patient looks reluctant. |
| "Why were you in the ER", "Vicodin" | Patient lowers head, looking ashamed. |

It is worth noting that these outputs not only inform the avatar's verbal replies but also trigger pre-defined gestures and facial expressions through keyword-based animation mapping. This configuration helps synchronizing the avatar's nonverbal behavior as the conversation flows. It also creates a cohesive feedback loop which connects the student's input, the API's reasoning process, and the avatar's multimodal presentation. Collectively, this architecture enhances both the authenticity and pedagogical value of the simulation.

### 4.4 Virtual Environment Engine

**4.4.1 Scene 1: Emergency Department.** In this scene (Fig. 4), the surrounding environment includes several medical devices mounted on the wall, an ECG monitor, a blood pressure device, a call button placed beside the bed, and a small wall-mounted speaker, all in place to mirror an authentic emergency room setting. The patient avatar is lying on the bed and looking ready to engage in conversations. She occasionally turns her head and eyes toward the user or nearby objects. At times, she raises her hand to touch her head or leg, adding a touch of realism to the encounter.

Users are positioned in a viewpoint facing directly to the patient avatar. During conversation, a panel appears in the lower-left corner of the screen and displays both the user inquiries and the patient responses in real-time as transcribed text. As shown in Fig. 4 and Fig. 5, the user's on-screen name is set as "Healthcare Provider" and the response from the patient avatar is set as "Jane Ryan". Users may choose to speak or type their inquiries, providing flexibility in interaction modes.

Nonverbal feedback is coordinated with dialogue through the keyword-triggered animation layer, as explained in Section 4.2. Five distinct response animations (see Table 1) were created in addition to the idle state, providing visual cues that complement spoken responses. For example, when the avatar's response is "no" or in other forms of rejection, a head-shake animation is synchronized with the verbal output. Similarly, when keywords such as "symptoms" or "discharge" are detected, the patient avatar lowers her head to convey embarrassment.

**4.4.2 Scene 2: Primary Care Office.** The second scene, as displayed in Fig. 5, takes place in a physician's office designed to convey a clinical, professional atmosphere. Main medical objects embedded in this scene include a blue examination table where the patient is seated and a medical lamp. The patient avatar maintains a neutral expression and looks directly toward the user. While sitting upright on the edge of the examination table, she occasionally looks around to scan the room and touches her head, mimicking subtle but context appropriate nonverbal cues.

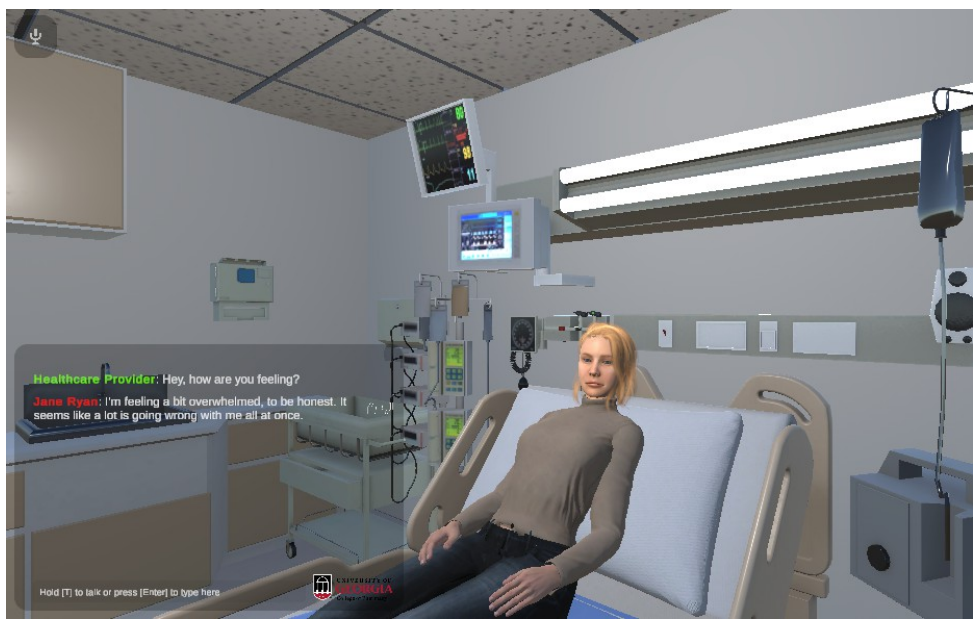

**Fig. 4.** Formal interaction in Scene 1 the emergency department

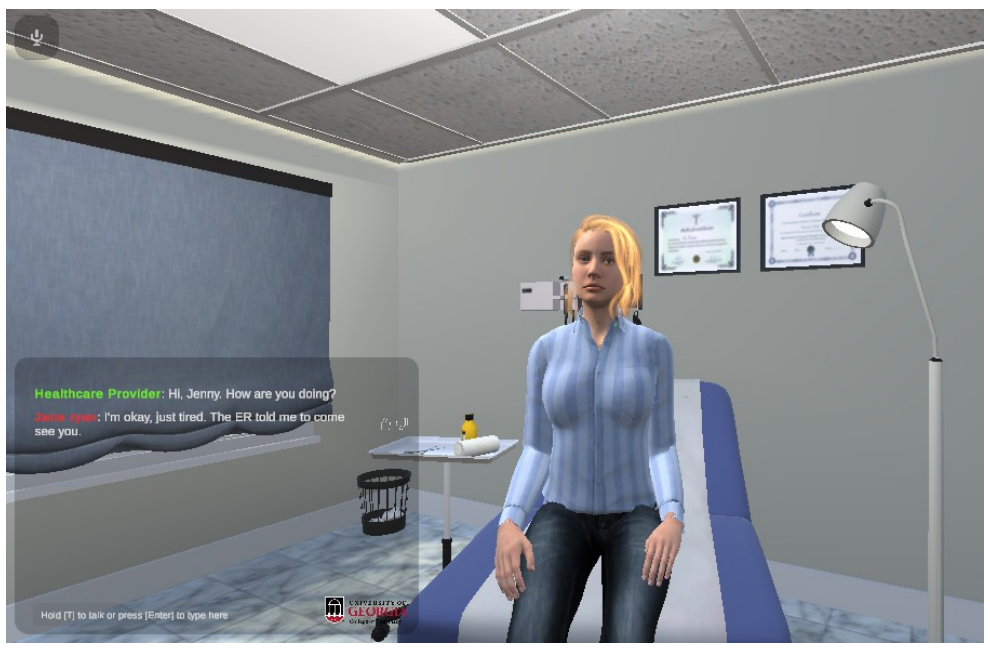

**Fig. 5.** Formal interaction in Scene 2 the primary care office

The keyword-to-animation mapping for the avatar's nonverbal responses in Scene 2 differs from Scene 1. Scene 2 contains fewer sets of trigger words and corresponding animations because the patient avatar has returned to an environment she's familiar with and her acute symptoms have been relieved, thanks to the healthcare team at the ED. Table 2 provides an overview of the user prompts and corresponding character animations they trigger, demonstrating how the system is designed to respond to scene-specific conversational cues with non-verbal expressions. The mechanism is identical to Scene 1. The Multimodal AI Engine detects keywords in user inputs, transitions the patient avatar to the mapped animation state, and simultaneously generates the verbal response.

### 4.5 Integration in Unity

Unity serves as the central integration platform that hosts both clinical scenarios and coordinates the three working engines. The patient avatar for both scenes along with animation clips were produced using the Character Creation Engine, specifically, Reallusion CC and iClone, and imported into Unity as assets. All speech and text inputs from users are routed and processed through the API by the Multimodal AI Engine, which is configured and trained contextually using authentic clinical scripts. The Virtual Environment Engine manages avatar rendering, generates real-time conversational response, and prompts synchronized, appropriate non-verbal responses (i.e., gestures, facial expressions, lip syncs) based on API output. Collectively, this architecture ensures all three engines operate within a single and continuous feedback loop: user input → API processing → real-time response → synchronized avatar animation, to support coherent, multimodal interactions across both clinical scenarios.

### 4.6 Usability Evaluation

**4.6.1 Data Collection.** A user testing session was conducted to evaluate the usability of the virtual simulation. To ensure a focused evaluation, four students with varying medical backgrounds were recruited as representative end users, as the actual IPE cohort will consist both undergraduate and graduate students. In line with Nielsen's usability guidelines [29], testing with three to five representative users is sufficient to uncover the majority of usability problems, as most issues tend to recur across users.

The session was intentionally designed to closely mimic how the IPE event will be structured. Each participant had general familiarity with IPE principles and practices, and was assigned a professional role (physician, pharmacist, nurse practitioner, and social worker). Participants interacted with the virtual patient under the guidance of a facilitator to collect relevant medical and psychosocial information. The evaluation goal focused on assessing the responsiveness, timeliness, and appropriateness of the virtual patient's responses, as well as identifying any usability issues that could hinder or impede the overall experience.

Throughout the session, user interactions were recorded by one primary facilitator and four observers who took detailed field notes. These notes focused on technical performance, user confusion, interaction breakdowns, and behavioral cues indicating difficulty, confusion, or hesitation.

**4.6.2 Data Analysis.** Two trained graduate students analyzed the observation notes collaboratively, and usability issues were classified using Nielsen's Usability Severity Rating Scale, ranging from 1 (Cosmetic) to 4 (Catastrophic). When discrepancies arose in coding or interpretation, the students consulted a senior professor with expertise in usability research and human-computer interaction to reach consensus. Findings were then synthesized to inform iterative improvements to the simulation's interface, dialogue structure, and instructional flow.

### 4.7 Ethical Approval

The usability evaluation was conducted under the broader institutional IRB protocol approved for the annual IPE event. The testing of the developed simulation was implemented as part of a quality improvement (QI) initiative aimed at enhancing instructional design and delivery.

## 5 Findings

### 5.1 Usability Results

A total of 22 usability issues were identified across both scenes (Fig. 6), which impacted users' interaction quality and task completion. Catastrophic and major usability issues were reported in themes as follows.

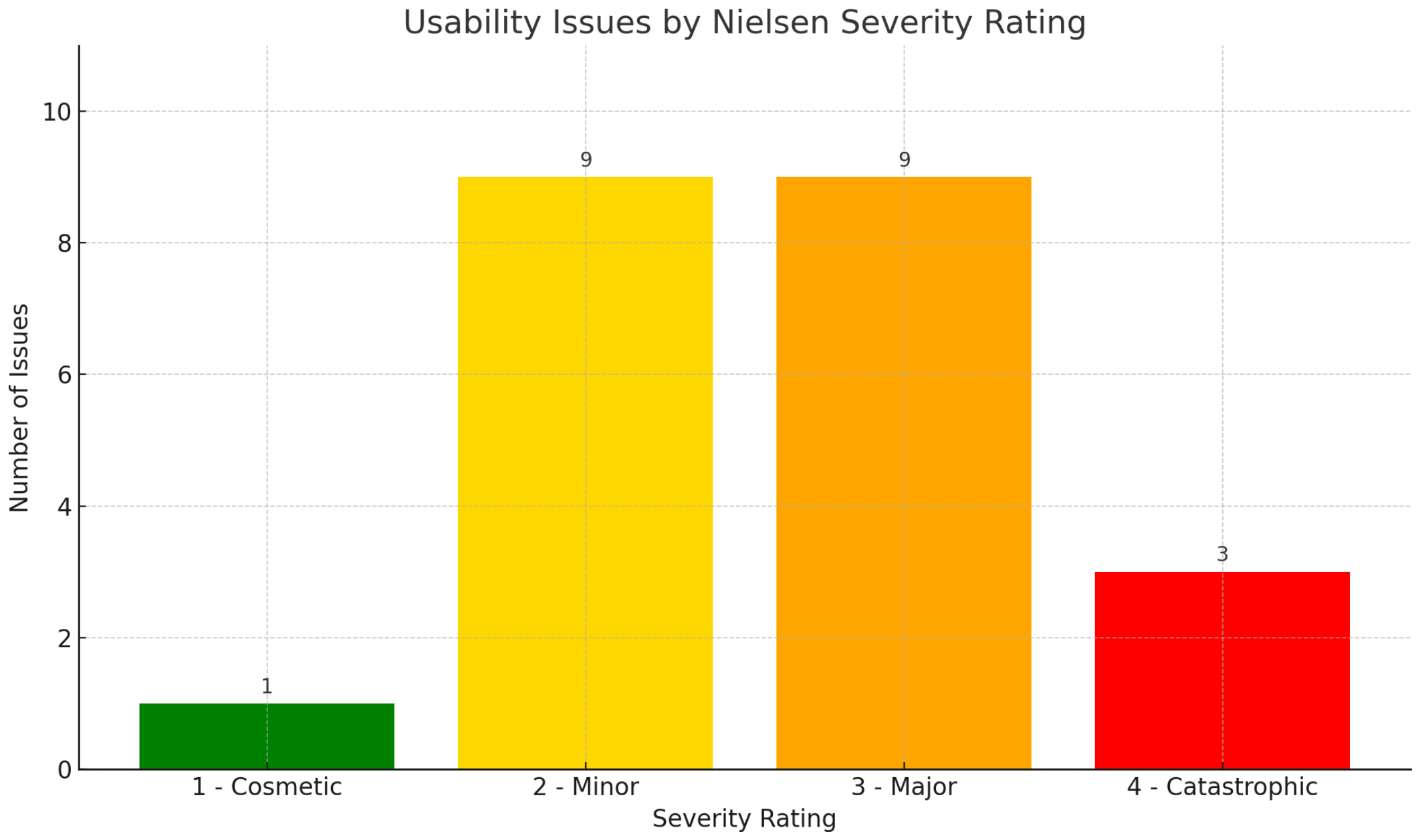

**Fig. 6.** Usability Issues By Nielsen's Severity Rating Scale

**5.1.1 Catastrophic Usability Issues (Severity 4, N=3).** Three catastrophic issues hindered usability during the test, all related to audio input and speech recognition failures. Participants experienced difficulty with the system randomly picking up out-of-context audio, even when the facilitator strictly followed the prescribed interaction protocol–pressing the "T" key to talk. After the facilitator released the "T" button and began instructing the participant on next steps, the system mistakenly registered the instruction as user input and responded accordingly, which significantly disrupted the flow of the simulation. This issue was classified as catastrophic, as it prevented task completion and compromised the educational experience.

**5.1.2 Major Usability Issues (Severity 3, N=9).** Three major usability issues were related to desynchronization and dialogue management errors. In some cases, the patient's facial animations and lip movements continued for up to 10 seconds after the audio response had stopped. This noticeable mismatch between speech and visual cues could create awkward silences and stalling the natural flow of interaction. In other cases, the patient cut off the participant and began responding before the participant had finished talking. It was also observed that the virtual patient repeated parts of its earlier responses, suggesting weak dialogue management logic. Taken together, these issues disrupted conversational flow and hindered the authenticity of the scenario, particularly in time-sensitive or emotional charged contexts that require fluid, empathetic communication.

Four issues were related to microphone control problems. The facilitator-controlled "T" button mechanism created significant usability challenges. The facilitator had to constantly watch student faces to determine when to activate and deactivate the microphone, creating high cognitive load and divided attention. This system also malfunctioned during the session when the microphone began "picking up everything being spoken without the T button being pressed", causing loss of turn control and crosstalk. Additionally, audio dropped out during avatar speaking, leaving participants confused about when they should continue. Students lacked clear non-verbal signals to indicate readiness to speak, forcing reliance on the facilitator's interpretation of facial cues.

A third theme emerged around instructional clarity and training gaps. Specifically, participants expressed confusion about the learning scenario itself. One student asked, "What is our diagnosis? What are we talking about?", indicating unclear learning objectives and case context. Post-session debriefs within the team identified the need for structured training blocks to familiarize users with the system before scenario participation.

Findings from the user testing session highlight the need for revisions and improvements in speech recognition system, real-time interaction design, and embodied feedback mechanisms to facilitate realistic, scalable training experiences for interprofessional education students. Major usability issues were addressed and the prototype was prepared for a second user testing session.

### 5.2 Design Revisions

Out of the 22 usability issues identified, all major and catastrophic (N=12) issues were resolved, which makes the prototype ready for the next round of testing. For example, to address the audio-routing failures that caused speech recognition failures, participants were instructed to install a lightweight audio mixing software that manages input routing and stabilizes streaming. The software allows users to control and lock microphone inputs across applications. This solution successfully reduced the microphone instability and largely improved the reliability of speech pipeline during an internal testing within the development team.

To address the desynchronization between verbal responses and lip movements, we recreated all animation clips. This is because during the initial design, we deliberately included approximately 100 frames at the beginning of each animation clip to ensure a smooth transition from the idle state to a specific state. However, these empty frames resulted in the noticeable visual latency from the user's perspective. To fix this, for each new animation clip, we deleted these empty frames to reduce transition time and achieve audiovisual synchronization. This adjustment led to a more natural and responsive character performance during user interaction.

To address the issue of instructional clarity and training gaps, although not tied directly to interface design, our team planned a brief orientation at the beginning of the IPE event. The presentation introduced the virtual simulation, outlined the learning objectives, and clarified role-specific responsibilities for all participating students. Additionally, facilitators were asked to guide students through a few rounds of low-stakes interactions with the virtual avatar to test audio functionality and internet connection, while setting up expectations for the scenarios. These procedural enhancements aim to reduce confusion, build user confidence, and promote a smoother, more focused training experience.

## 6 Conclusion

### 6.1 Discussion and Future Directions

This paper details the development and initial evaluation of AIMS, an AI-enhanced screen-based virtual simulation designed for IPE. While the current version is still considered a work-in-progress, it demonstrates technological innovations in three key areas: (1) it leverages multimodal LLMs for role-based, contextually grounded verbal interactions; (2) it implements a character rendering pipeline that generates dynamic, sustained non-verbal cues; and (3) it operates through a web-based platform that improves accessibility. AIMS' technical capabilities and accessibility features directly address two gaps identified in our review: the limited integration of real-time conversational AI in virtual patient systems and the challenge of scaling virtual simulations for large-scale deployment. Compared to immersive VR, screen-based simulations offer lower levels of presence and embodied realism, which may affect experiential learning. However, such simulations are more accessible, cost-efficient, and scalable especially for the IPE event in this context where users are geographically distributed.

The significance and impact of AIMS lie in showing what becomes possible when LLM reasoning, embodied animations, and scalable virtual environments are purposefully integrated around well-defined learning objectives. It can function both as an instructional platform and as an experimental space for investigating AI-facilitated clinical education. Our next steps will focus on systematic refinement and evaluation of AIMS in preparation for the large-scale implementation during the annual IPE event hosted at our home institution, which will involve over 400 students and faculty facilitators. We will measure perceptions using AIMS and self-reported competencies after the event. Specifically, we aim to investigate: (1) To what extent do healthcare provider students across health professions perceive AIMS as usable, accessible, and relevant for interprofessional training? and (2) How does participation in the AIMS-supported IPE event impact students' self-reported competencies in collaborative communication and team-based decision-making? These questions will guide future data collection using post-activity surveys, usability questionnaires, and focus group interviews.

### 6.2 Limitations

This work details an iterative, formative design and preliminary evaluation case of AIMS, an AI-enhanced, screen-based virtual simulation. However, findings should be interpreted in light of several limitations. First, the system was implemented on a web platform, although highly accessible, does not provide the full sensory immersion that traditional VR environments offer. This trade-off may limit embodied presence and certain nonverbal communication affordances. Second, the initial usability testing involved a small group of four representative users, although sufficient to identify major design issues, does not support claims about broader system effectiveness. Last but not least, the current work focused on system usability not educational effectiveness or learning outcomes. Future work will examine learning outcomes at scale following the IPE event.

## Acknowledgment

This study was supported by the Learning Technology Grant sponsored from the Center for Teaching and Learning at the University of Georgia.